\newcommand{\beq}{  \begin{eqnarray}}
\newcommand{\eeq}{  \end{eqnarray}}

\documentclass[prl,twocolumn,showpacs,preprintnumbers,amsmath,amssymb]{revtex4-1}
\usepackage{subfig}
\usepackage{graphicx}
\usepackage{dcolumn}
\usepackage{bm}

\begin{document}
    	
\title{Comment on ``Measurement of a Large Chemical Reaction Rate between Ultracold Closed-Shell $^{40}$Ca Atoms and Open-Shell $^{174}$ Yb$^{+}$ 
Ions Held in a Hybrid Atom-Ion Trap"}
    
\author{B. Zygelman }\email{bernard@physics.unlv.edu}
\author{Robert Hunt}
\affiliation{%
Department of Physics and Astronomy, University of Nevada, Las Vegas, Las Vegas NV 89154
}%

\date{\today}
\begin{abstract}
 \end{abstract}

\maketitle
In a recent Letter, Rellergert et al.\cite{reller11} presented the results of measurements of the chemical reaction rate
for $^{40}$Ca atoms to recombine with $^{174}$ Yb$^{+}$ ions in a hybrid atom-ion trap. They obtained the rate K= $2 \times 10^{-10}$ cm$^{3}$ s$^{-1}$
for temperatures that range from 1 mK to 10 K and which they attributed to losses incurred by radiative association (RA) and radiative charge transfer (RCT).
In this Comment we calculate the radiative transition rate coefficient K, and find that it underestimates the measured rate by about five orders of magnitude. 
We propose possible explanations for the large
discrepancy. We use atomic units below.
 
An upper limit for the total RA+RCT rate can be obtained using the local optical potential method\cite{zyg89} in which
the total cross section for radiative relaxation of the $^{40}$Ca,  $^{174}$ Yb$^{+}$ pair is given by
\beq
\sigma=\frac{\pi}{k^{2}} \sum_{J} (2 J+1) \Bigl (1 -\exp(-4 \eta_{J}) \Bigr ).
\label{1.1}
\eeq
Here $k=\sqrt{2 \mu E}$ is the wavenumber corresponding to a collision energy $E$, $\eta_{J}$ is the
imaginary part of the phase shift for the radial partial wave amplitude $f_{J}(r)$ which obeys,
\beq
\frac{d^{2} f_{J}}{d r^{2}} -\frac{J(J+1)}{r^{2}} f_{J} -2\mu [V(r)-i A(r)/2] f_{J} + k^{2} f_{J} = 0. \nonumber 
\eeq
$V(r)$ is the potential energy curve for the $A^{2}\Sigma^{+}$ state of the molecular ion,
$J$ is an angular momentum, and
$
A(r) = \frac{4}{3} D^{2}(r) \frac{|\Delta E(r)|^{3}}{c^{3}}
$
is the radiative transition rate.
Here $D(r)$ is the transition dipole moment for the  $A^{2}\Sigma^{+}$ and  $X^{2}\Sigma^{+}$ states of the molecular ion,
$\Delta E$ is the energy defect between them, $c=137$ is the speed of light, and $\mu$ is the system reduced
mass.
\begin{figure}[ht]
\centering
\includegraphics[scale=0.4]{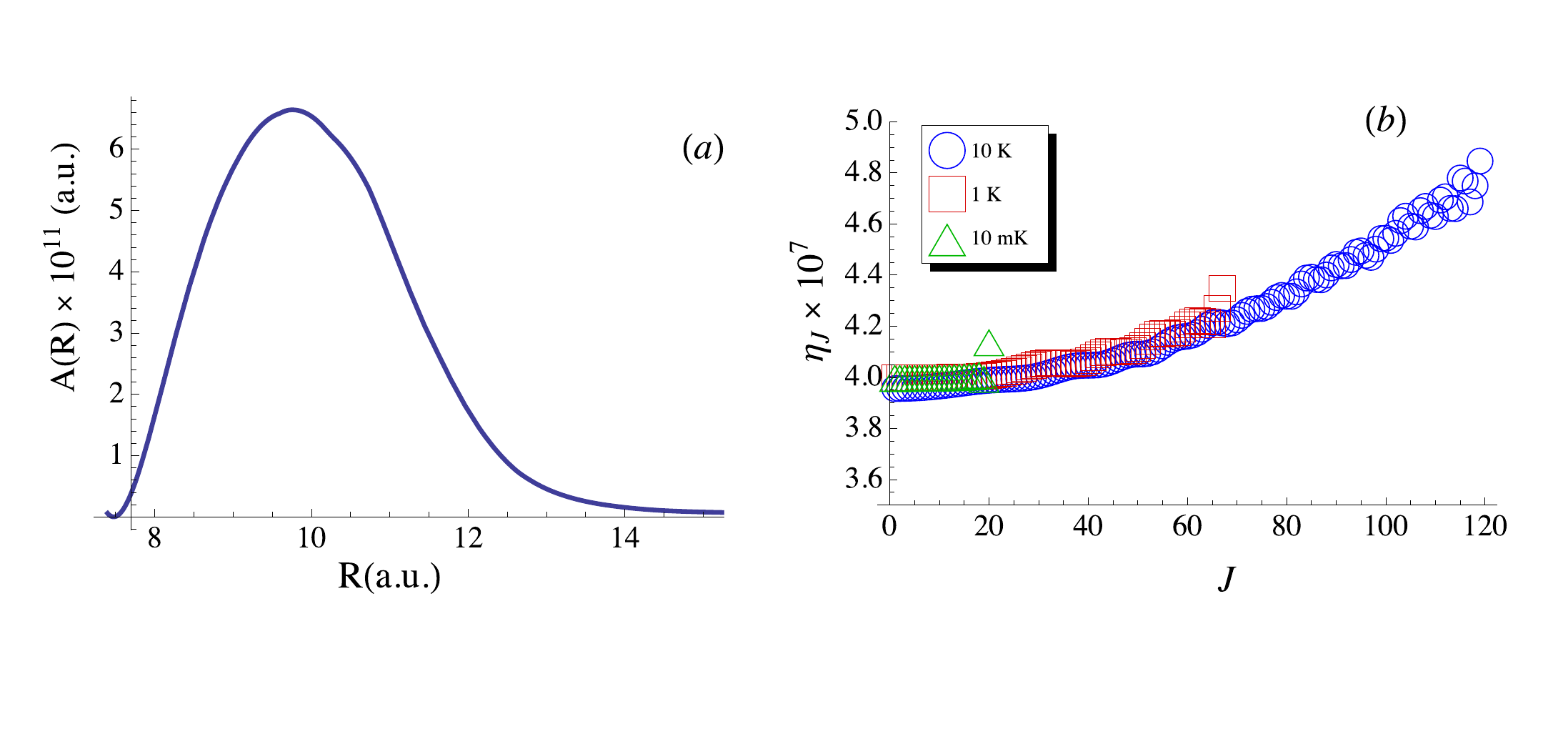}
\caption{\label{fig:fig1}(Color online) (a) Plot of $A(r)$. (b) Calculated values of $\eta_{J}$ at temperatures T=10,1 and 0.01 Kelvin.}
\end{figure}
The molecular parameters needed to construct
$A(r)$, shown in Fig. (1a), where taken from Figs. (3b,c) in Ref.\cite{reller11} and, at large $r$, are fitted to the polarization potentials. 

In our calculations we found that $ \eta_{J}$, shown in Fig. (1b), varies slowly for all $J < J_{max}$ at the collision energies of interest. For $J$
greater than
 $ J_{max} \approx \sqrt[4]{8 \mu k^{2} C_{4}}$ 
classical trajectories of the atom-ion pair are prevented from penetrating the centrifugal barrier. Using the calculated 
values for $ \eta_{J}$ we obtain,
\beq
&& \sigma \approx \frac{4 \pi \langle \eta \rangle J_{max}^{2}}{k^{2}} = \frac{4 \pi \langle \eta \rangle \sqrt{8 \mu C_{4}}} {k} \nonumber \\
&& {\rm K} \approx \langle v \rangle \sigma = 4 \pi \langle \eta \rangle \sqrt{\frac{8 C_{4}}{\mu}}
\label{1.3}
\eeq  
where $\langle \eta \rangle $ is an average of $ \eta_{J}$ which we take here to have the approximate value $4.3 \times 10^{-7}$. 
$\langle v\rangle$ is the average collision speed, and
using $C_{4}=78.5$\cite{reller11}
we obtain  K $ \approx 3.4 \times 10^{-15} $ cm$^{3}$ s$^{-1}$, which is about 5 orders of magnitude smaller than the measured rate. 

Possible reasons for this discrepancy are:
(a) A profound breakdown in the local optical potential approximation, which follows from the standard theory e.g. see Ref. \cite{zyg88}.
 If that is the case, the results of studies that employed it need to be re-evaluated.
(b) The neglect of shape resonances. For $J>J_{max}$ quantum barrier tunneling can lead to large cross sections that dominate
contributions to the rate coefficient at low temperatures\cite{zyg98}. 
(c) Uncertainty in the molecular data, and/or the neglect of non-adiabatic effects.
Of those choices we believe items (b,c) to be the most plausible. In either case, our studies suggest that additional theoretical
and laboratory work is required to resolve the issues raised.

\end{document}